\documentclass[conference]{IEEEtran}
\IEEEoverridecommandlockouts
\usepackage{cite}
\usepackage{amsmath,amssymb,amsfonts}
\usepackage{glossaries}
\newacronym{asic}{ASIC}{Application-Specific Integrated Circuit}
\newacronym{awgn}{AWGN}{Additive White Gaussian Noise}

\newacronym{bm3d}{BM3D}{Block-Matching 3D}

\newacronym{cnn}{CNN}{Convolutional Neural Network}
\newacronym{crt}{CRT}{Cathode Ray Tube}
\newacronym{cpu}{CPU}{Central Processing Unit}
\newacronym{cr}{CR}{Character Recognition}

\newacronym{dvi}{DVI}{Digital Visual Interface}
\newacronym{dncnn}{DnCNN}{Denoising Convolutional Neural Network}
\newacronym{dnn}{DNN}{Deep Neural Network}
\newacronym{dp}{DP}{Display Port}
\newacronym{dl}{DL}{Deep Learning}

\newacronym{em}{EM}{Electro Magnetic}
\newacronym{emsec}{EMSEC}{Emission Security}
\newacronym{xai}{XAI}{eXplainable Articial Intelligence}

\newacronym{fcn}{FCN}{Fully Convolutional Network}
\newacronym{fc}{FC}{Fully Connected}
\newacronym{fpga}{FPGA}{Field Programmable Gate Array}
\newacronym{fpn}{FPN}{Fixed Pattern Noise}

\newacronym{gcbd}{GCBD}{GAN-CNN based Blind Denoiser}

\newacronym{gpu}{GPU}{Graphics Processing Unit}
\newacronym{gan}{GAN}{Generative Adversarial Networks}

\newacronym{g2g}{G2G}{Generated-Artificial-Noise to Generated-Artificial-Noise}

\newacronym{hdmi}{HDMI}{High-Definition Multimedia Interface}

\newacronym{ipe}{IPE}{Information processing equipment}
\newacronym{ilsvrc}{ILSVRC}{ImageNet Large Scale Visual Recognition Competition}

\newacronym{lcd}{LCD}{Liquid Crystal Display}
\newacronym{lvds}{LVDS}{Low-Voltage Differential Signaling}

\newacronym{mse}{MSE}{Mean Square Error}
\newacronym{mae}{MAE}{Mean Absolute Error}
\newacronym{mwcnn}{MWCNN}{Multi-level Wavelet Convolutionnal Neural Network}

\newacronym{n2n}{N2N}{Noise2Noise}
\newacronym{n2v}{N2V}{Noise2Void}
\newacronym{n2s}{N2S}{Noise2Self}

\newacronym{ocr}{OCR}{Optical Character Recognition}

\newacronym{psnr}{PSNR}{Peak Signal to Noise Ratio}

\newacronym{red}{RED}{Residual Encoder-Decoder networks}
\newacronym{roi}{RoI}{Region of Interest}
\newacronym{rois}{RoIs}{Regions of Interest}
\newacronym{rf}{RF}{Radio Frequency}
\newacronym{rmse}{RMSE}{Root Mean Square Error}
\newacronym{rpn}{RPN}{Region Proposal Network}
\newacronym{rnn}{RNN}{Recurrent Neural Network}

\newacronym{sdr}{SDR}{Software-Defined Radio}
\newacronym{sgn}{SGN}{Self-Guided Network}
\newacronym{snr}{SNR}{Signal to Noise Ratio}
\newacronym{srresnet}{SRResNet}{Super-Resolution Residual Network}
\newacronym{ssim}{SSIM}{Structure Similarity}

\newacronym{vga}{VGA}{Video Graphics Array}

\usepackage{subfig}
\usepackage{multirow}
\usepackage{algorithmic}
\usepackage{graphicx}
\usepackage{textcomp}
\usepackage{xcolor}
\usepackage{hhline}
\makeatletter

\def\ps@IEEEtitlepagestyle{
  \def\@oddfoot{\mycopyrightnotice}
  \def\@evenfoot{}
}
\def\mycopyrightnotice{
  {978-1-7281-9320-5/20/\$31.00 ©2020 IEEE} 
  \gdef\mycopyrightnotice{}
}

\@ifundefined{showcaptionsetup}{}{
 \PassOptionsToPackage{caption=false}{subfig}}
\usepackage{subfig}
\makeatother

\usepackage{eso-pic}

\definecolor{myred}{rgb}{0.8745,0.1961,0.1412}

\definecolor{myblue}{rgb}{0,0,1}

\def\BibTeX{{\rm B\kern-.05em{\sc i\kern-.025em b}\kern-.08em
    T\kern-.1667em\lower.7ex\hbox{E}\kern-.125emX}}
\begin{document}

\title{NoiseBreaker: Gradual Image Denoising Guided by Noise Analysis
\thanks{This work is supported by the "Pôle d'Excellence Cyber".}
}

\author{\IEEEauthorblockN{Florian Lemarchand$^{\star}$ \qquad Thomas Findeli$^{\star}$ \qquad  Erwan Nogues$^{\star\dagger}$ \qquad Maxime Pelcat$^{\star}$}
\IEEEauthorblockA{$^{\star}$ Univ. Rennes, INSA Rennes, IETR - UMR CNRS 6164, France \\
$^{\dagger}$ DGA-MI, Bruz, France \\
Emails: firstname.lastname@insa-rennes.fr\\}}

\maketitle

\begin{abstract}
   
   Fully supervised deep-learning based denoisers are currently the most performing image denoising solutions. 
   However, they require clean reference images. 
   When the target noise is complex, e.g. composed of an unknown mixture of primary noises with unknown intensity, fully supervised solutions are hindered by the difficulty to build a suited training set for the problem. 
   
   This paper proposes a gradual denoising strategy called NoiseBreaker that iteratively detects the dominating noise in an image, and removes it using a tailored denoiser. The method is shown to strongly outperform state of the art blind denoisers on mixture noises. Moreover, noise analysis is demonstrated to guide denoisers efficiently not only on noise type, but also on noise intensity. NoiseBreaker provides an insight on the nature of the encountered noise, and it makes it possible to update an existing denoiser with novel noise profiles. This feature makes the method adaptive to varied denoising cases.  

\end{abstract}


\glsresetall

\section{Introduction}\label{sec:intro}

Image denoising, as a sub-domain of image restoration, is an extensively studied problem~\cite{buades_review_2005} though not yet a solved one~\cite{chatterjee_is_2010}. The objective of a denoiser is to generate a \emph{denoised} image $\hat{x}$ from an observation $y$ considered a \emph{noisy} or corrupted version of an original \emph{clean} image $x$. $y$ is generated by a noise function $h$ such that $y = h(x)$. 
A vast collection of noise models exist~\cite{boyat_review_2015} to represent $h$. Examples of frequently used models are Gaussian, Poisson, Bernoulli, speckle or uniform noises. 
While denoisers are constantly progressing in terms of noise elimination level~\cite{dabov_image_2007,zhang_beyond_2017,liu_multi-level_2018}, most of the published techniques are tailored for and evaluated on a given \textit{primary} noise distribution (i.e. respecting a known 
distribution). They exploit probabilistic properties, of the noise they are specialised for, to distinguish the noise from the signal of interest. 

Complex noises are more application specific than primary noises, but their removal constitutes an identified and important issue. Real photograph noise~\cite{plotz_benchmarking_2017} is for instance a sequential composition of primary noises~\cite{gow_comprehensive_2007}, generated by image sensor defects. Images retrieved from electromagnetic side channel attacks of screen signal~\cite{lemarchand_electro-magnetic_2020} also contain strong noise compositions, as the interception successively introduces several primary noises. The distributions of these real-world noises can be approached using noise composition models, also called \textit{mixtures}. Mixture noise removal has been less extensively addressed in the literature than primary noise removal. When modeling experimental noises, different types of mixtures can be used. The corruption can be \emph{spatially} composed such that each pixel $\rho$ of an image can be corrupted by a specific distribution $\eta(\rho)$. $h$ is then composed of the set $\{\eta(\rho), \rho\in x\}$. This type of noise is called a \textit{spatial mixture}~\cite{cha_gan2gan:_2019, batson_noise2self:_2019}. The mixture noise can also be considered as the result of $n$ primary noises applied with distributions $\eta_i, i\in\{0..n-1\}$ to each pixel $\rho$ of the image $x$. This type of noise is called a \textit{sequential mixture}. This paper focuses on sequential mixture noises removal.

Real-world noises, when not generated by a precisely known random process, are difficult to restore with a discriminative denoiser that requires $(y,x)$ pairs of observed and clean images. The lack of clean images may cause impossibility to build large supervised databases. \textit{Blind denoising} addresses this lack of supervised dataset and consists of learning denoising strategies without exploiting clean data. Even if they solve the clean data availability issues, blind denoisers do not provide any hint on the types of noise they have encountered. 
Our approach takes the opposite side and focuses on efficient noise analysis.

Knowing the types of noise in an image has several advantages. First, it enables to identify primary noises and leverages a deep understanding of their individual removal. Secondly, by decomposing the mixture denoising problem into primary ones, a library of standard denoisers can be built to answer any noise removal problem. This second point is developed as the central point of our proposed method. Thirdly, a description of the image noise content helps identify the physical source of data corruption. Finally, under the assumption of sequential noise composition, it is possible to build a denoising pipeline from the identified noise distribution. The noise distribution being known, large training databases generation becomes feasible.

In this paper, an image restoration method called NoiseBreaker (NBreaker) is proposed. It recursively detects the dominating noise type in an image as well as its noise level, and removes the corresponding noise. 
This method assumes that the target noise is a sequential mixture of known noises. The resulting step-by-step gradual strategy is driven by a noise analysis of the image to be restored. The solution leverages a set of denoisers trained for primary simple noise distribution and applied sequentially following the prediction of a noise classifier.

The manuscript is organized as follows. Section~\ref{sec:related_work} presents related work on image noise analysis and image denoising. Section~\ref{sec:contrib} details the proposed solution. Section~\ref{sec:expe} evaluates the proposal on synthetic mixture noise and situates among close state of the art solutions. Section~\ref{sec:conclu} concludes the paper and gives future perspectives.

\section{Related Work}\label{sec:related_work}
Fully supervised deep learning denoisers forge a restoration model from pairs of \textit{clean} and corresponding \textit{noisy} images. Following the premises of \gls{cnn} based denoisers~\cite{jain_natural_2009}, different strategies have been proposed such as residual learning~\cite{zhang_beyond_2017, ledig_photo-realistic_2017}, skip connections~\cite{mao_image_2016}, the use of transform domain~\cite{liu_multi-level_2018} or self-guidance for fast denoising~\cite{gu_self-guided_2019}. The weakness of supervised denoisers is their need of large databases, including clean images.

To overcome this limitation, different weakly supervised denoisers have been proposed.
In particular, \textit{blind} denoisers are capable of removing noise without clean data as reference. First studies on blind denoising aimed at determining the level of a known noise in order to apply an adapted human-expert based denoising (e.g. filtering). \gls{n2n}~\cite{lehtinen_noise2noise:_2018} has pioneered learning-based blind denoising using only noisy data. It demonstrates the feasibility of learning a discriminative denoiser from only a pair of images representing two independent realisations of the noise to be removed. Inspired by the latter, \gls{n2v} is a recent strategy that trains a denoiser from only the image to be denoised.

Most denoisers are designed for a given primary noise. The most addressed distribution is \gls{awgn}~\cite{mao_image_2016, vincent_stacked_2010,  jin_flexible_2019}. To challenge denoisers and approach real-world cases, different types of mixture noises have been proposed in the literature. They are all created from the same few primary noises. A typical example of a spatially composed mixture noise~\cite{zhao_robust_2014} is made of 10\% of uniform noise $[-s, s]$, 20\% of Gaussian noise $\mathcal{N}(0,\sigma_0)$ and 70\% of Gaussian noise $\mathcal{N}(0,\sigma_1)$. These percentages refer to the amount of pixels, in the image, corrupted by the given noise. This type of spatial mixture noise has been used in the experiments of \gls{g2g}~\cite{cha_gan2gan:_2019}. 
An example of sequential mixture noise is used to test the recent Noise2Self method~\cite{batson_noise2self:_2019}. It is composed of a combination of Poisson noise, Gaussian noise, and Bernoulli noise. In~\cite{lemarchand_opendenoising_2020}, authors compare methods designed for \gls{awgn} removal when trained and evaluated on more complex noises such as sequential mixtures of Gaussian and Bernoulli distributions. Experimental results show that denoising performances severely drop on complex noises even when using supervised learning methods like \gls{dncnn}. This observation motivates the current study and the chosen sequential mixture noise. Future studies will address spatial mixture noises.

In~\cite{schmidt_bayesian_2011}, authors propose a noise level estimation method integrated to a deblurring method. Inspired from the latter proposal, the authors of~\cite{liu_single-image_2013} estimate the standard deviation of a Gaussian distribution corrupting an image to apply the accordingly configured \gls{bm3d} filtering~\cite{dabov_image_2007}. This can be interpreted as a noise characterization, used to set parameters of a following dedicated denoising. To the best of our knowledge, the present study is the first to use noise type and intensity classification for denoising purposes.

The image classification domain has been drastically modified by deep learning methods since LeNet-5~\cite{lecun_gradient-based_1998}.
With the development of tailored algorithms and hardware resources, deeper and more sophisticated neural networks are emerging. 
Seeking a good trade-off between classification efficiency and hardware resources, MobileNets~\cite{howard_mobilenets:_2017} is a particularly versatile family of classifiers. MobileNetV2 has become a standard for resource aware classification. In our study, we use MobileNetV2 pre-trained on ImageNet and fine-tuned for noise classification~\cite{tajbakhsh_convolutional_2016}.

Recent studies have proposed classification-based solutions to the image denoising problem \cite{sil_convolutional_2019, liu_classification_2020}. Sil et al.~\cite{sil_convolutional_2019} address blind denoising by choosing one primary noise denoiser out of a pool, based on a classification result. NoiseBreaker goes further by considering composition noises, sequentially extracted from the noisy image using a sequence of classify/denoise phases. In~\cite{liu_classification_2020}, authors adopt a strategy close to NoiseBreaker. However, NoiseBreaker differentiates from that proposal by refining the classes into small ranges of primary noises. We demonstrate in the results Section~\ref{sec:expe} that NoiseBreaker outperforms \cite{liu_classification_2020}. 
The main reason for NoiseBreaker to outperform the results of~\cite{liu_classification_2020} is that the NoiseBreaker denoising pipeline is less constrained. As an example, when a Gaussian noise is detected by the classifier of \cite{liu_classification_2020}, the first denoising step is always achieved using a Gaussian denoiser. Our proposal does not compel the denoising process to follow a predefined order and lets the classifier fully decide on the denoising strategy to be conducted.

In this paper, we propose to tackle the denoising of sequential mixture noises via an iterative and joint classification/denoising strategy. Our solution goes further than previous works by separating the denoising problem into simpler steps optimized separately.

\section{Gradual Denoising Guided by Noise Analysis}\label{sec:contrib}

Our proposed solution is named NoiseBreaker (NBreaker).
It is the combination of a noise classifier and several denoisers working alternatively to gradually restore an image.
NBreaker is qualified as gradual because it denoises the input image iteratively, alternating between noise detection and removal. 
To provide this step-by-step restoration, 
NBreaker leverages a classifier acting as a noise analyser and further guiding a pool of denoisers specialized to \textit{primary noise} distributions. Both the analyser and the gradual denoising strategy are detailed here-under. NBreaker is able to handle numerous mixture noises at inference time without information on the composition of the mixture and without previous contact with the mixture. Hence, NBreaker is a blind denoiser at inference time.

\subsection{Noise Analysis}\label{subsec:noise_analysis}
The objective of the noise classifier $\mathcal{C}$ is to separate images into $n$ noise classes. 
A noise class is defined by a noise type and a range of parameter values contained in its images. 
When for a noise type no parameter exist or an only range is used, the class is denoted using $\eta_i$. 
Otherwise, it is denoted using $\eta_{i,j}$ with $i$ an index among a list $H$ of noise types and $j$ an index for the different ranges of a given noise type. $\eta_i$ (or $\eta_{i,j}$) is said to be a \textit{primary} noise. 
Note that one class 
does not contain any noise and serves to identify \textit{clean} images. 

The architecture of the classifier is made of a feature extractor, called \textit{backbone}, followed by two \gls{fc} layers, said as \textit{head}. The feature extractor, responsible for extracting knowledge out of the images, is the one of MobileNetV2. 
The two \gls{fc} layers have respectively 1024 and $n$ units, where $n$ is the number of classes of the classification problem. The input size is chosen to be $224\times224$ as in the original MobileNetV2 implementation. The first \gls{fc} layer has ReLu activation while the second uses softmax activation to obtain outputs in $[0,1]$, seen as certainty coefficients. The output of this second \gls{fc} layer, passed through an argmax function, gives the label with the highest certainty coefficient.

\begin{table}
    \centering
    \begin{tabular}{c l l l l}
        Class & \multicolumn{1}{c}{Noise Type} & \multicolumn{1}{c}{Parameters} & \multicolumn{1}{c}{Denoiser} \\ 
        \hline        
        $\eta_{0,0}$ & \multirow{3}{*}{Gaussian ($\mathcal{N}$)}  & $\sigma_g=[0,15]$ & \multirow{3}{*}{MWCNN~\cite{liu_multi-level_2018}}\\ 
        $\eta_{0,1}$ &   & $ \sigma_g=]15,35]$  \\
        $\eta_{0,2}$ &   & $ \sigma_g=]35,55]$ \\
        \hline   
        $\eta_{1,0}$ & \multirow{3}{*}{Speckle ($\mathcal{S}$)} & $\sigma_s=[0,15]$ & \multirow{3}{*}{SGN~\cite{gu_self-guided_2019}} \\ 
        $\eta_{1,1}$ &   & $ \sigma_s=]15,35]$  \\
        $\eta_{1,2}$ &  & $ \sigma_s=]35,55]$ \\
        \hline   
        $\eta_{2,0}$ &  \multirow{2}{*}{Uniform ($\mathcal{U}$)}  & $ s=[-10,10]$ & \multirow{2}{*}{SGN~\cite{gu_self-guided_2019}} \\ 
        $\eta_{2,1}$ &    & $ s=[-50,50]$  \\
        \hline   
        $\eta_{3}$ & Bernoulli ($\mathcal{B}$) & \multicolumn{1}{l}{$p=[0,0.4]$} & SRResNet~\cite{ledig_photo-realistic_2017}\\ 
        \hline  
        $\eta_{4}$ & Poisson ($\mathcal{P}$) & \multicolumn{1}{c}{$\emptyset$} & SRResNet~\cite{ledig_photo-realistic_2017}\\
        \hline  
        $\eta_{5}$ & Clean ($\emptyset$) & \multicolumn{1}{c}{$\emptyset$} & \multicolumn{1}{c}{$\emptyset$} \\
    \end{tabular}
    \caption{List of classes for NBreaker, the noise they represent and the related denoiser. Gaussian and Speckle classes are zero mean. The denoisers are selected from a benchmark study.} 
    \label{tab:primary_denoisers_NBreaker_E}
\end{table}

\subsection{Gradual Denoising}\label{subsec:gradual_denoiser}

A noisy image is given to NBreaker. That image is fed to the classifier $\mathcal{C}$ trained to differentiate noise classes. $\mathcal{C}$ supplies a prediction $\eta_i$ to $\mathcal{G}$, the gradual denoising block. $\mathcal{G}$ selects the corresponding denoiser $\mathcal{D}(\eta_i)$. $\mathcal{D}(\eta_i)$ is said to be a \textit{primary denoiser} specialized for the denoising of $\eta_i$ noise class. A primary denoiser is a denoiser trained with pairs of clean and noisy images from the $\eta_i$ class. The process ($\mathcal{C}$ followed by $\mathcal{G}$) is iterative and loops $n$ times until $\mathcal{C}$ detects the class \textit{clean}. 

 \begin{figure*}
    \centering
    \includegraphics[width=0.95\linewidth]{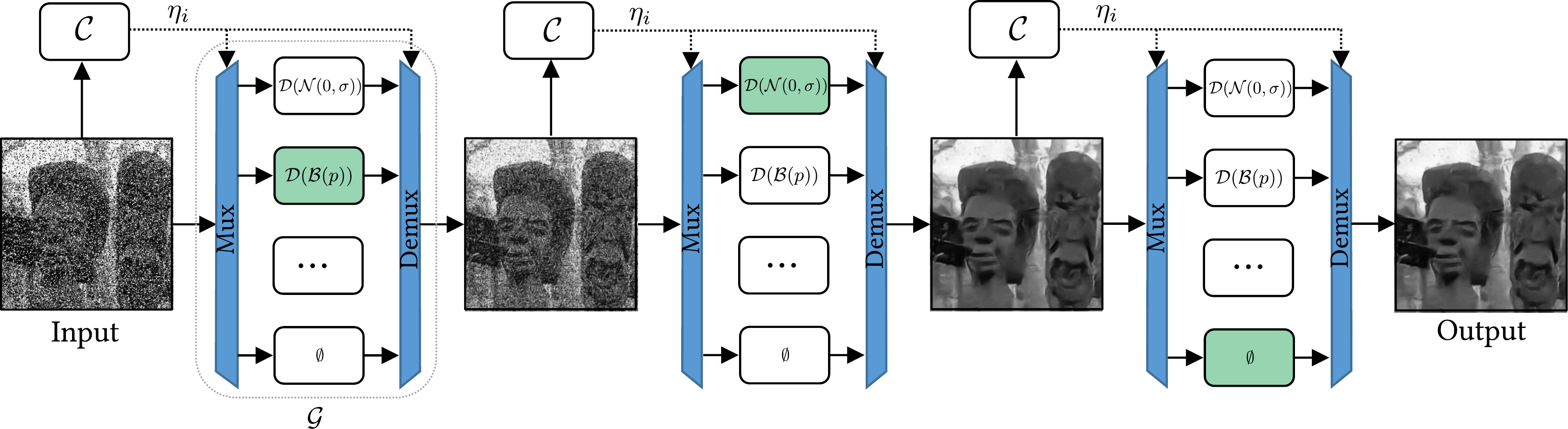}
\caption{Example of NBreaker gradual denoising. A noisy input image is fed to the classifier $\mathcal{C}$ which outputs a prediction $\eta_i$. This prediction drives the gradual denoising block $\mathcal{G}$ that selects the primary denoiser $\mathcal{D}(\eta_i)$ to be applied. The process runs for two steps until no noise is detected by $\mathcal{C}$.}
  \label{fig:mux} 
\end{figure*}

The architecture of the gradual denoiser depends on the classifier $\mathcal{C}$ since it drives the primary denoiser selection. 
An example of gradual denoising is given in Figure~\ref{fig:mux} where two noise classes are successively detected and treated.
The description of the classes considered for NBreaker is given by Table~\ref{tab:primary_denoisers_NBreaker_E}, note that Gaussian and Speckle classes are zero mean. 
For each class, several state of the art denoising architectures have been compared to be used as primary denoisers. The comparison was done using the OpenDenoising benchmark tool~\cite{lemarchand_opendenoising_2020}. The compared methods are \gls{mwcnn}~\cite{liu_multi-level_2018}, \gls{sgn}~\cite{gu_self-guided_2019}, \gls{srresnet}~\cite{ledig_photo-realistic_2017} and \gls{dncnn}~\cite{zhang_beyond_2017}. The results shown that results are consistent through a noise type and no gain is obtained choosing a different architecture for the noise levels of a noise type. From that benchmark study, \gls{mwcnn}~\cite{liu_multi-level_2018} is used for Gaussian noise type, \gls{sgn}~\cite{gu_self-guided_2019} for Speckle and uniform and \gls{srresnet}~\cite{ledig_photo-realistic_2017} for Bernoulli and Poisson. The noise types associated to their selected architecture are shown in Table~\ref{tab:primary_denoisers_NBreaker_E}.

\begin{table}
    \centering
    \begin{tabular}{c c c }
         & Noise 1 & Noise 2 \\ 
        \hline        
        $C_0$ & $ \mathcal{N}(0 ,[0,55])$ &  $\mathcal{B}([0,0.4])$\\ 
        \hline   
        $C_1$ & $ \mathcal{N}(0 ,[0,55])$ & $\mathcal{S}(0 ,[0,55])$   \\
        \hline   
        $C_2$ &  $ \mathcal{N}(0 ,[0,55])$ &   $\mathcal{P}$ \\ 
        \hline   
        $C_3$ & $\mathcal{B}([0,0.4])$ & $\mathcal{S}(0 ,[0,55])$  \\ 
        \hline  
        $C_4$ & $\mathcal{B}([0,0.4])$ & $\mathcal{P}$ \\
        \hline  
        $C_5$ & $\mathcal{S}(0 ,[0,55])$ & $\mathcal{P}$\\
    \end{tabular}
    \caption{Definition of the mixture noises used for evaluation. Noise~1 is applied on the sample followed by Noise~2.}
    \label{tab:test_mixtures_table}
\end{table}

\begin{table*}
    \centering
 
    \begin{tabular}{c l c c c c c c }
    \hhline{========}
        Dataset & \multicolumn{1}{c}{Denoiser} & $C_0$ & $C_1$ & $C_2$ & $C_3$ & $C_4$ & $C_5$\\
    \hline

        \multirow{5}{*}{BSD68 Grayscale} & Noisy& 12.09/0.19 & 16.98/0.36 & 18.21/0.42 & 14.05/0.28 & 13.21/0.24 & 24.96/0.73 \\

        & BM3D~\cite{dabov_image_2007} & 21.49/0.54 & 24.00/0.61 & 24.28/0.62 & 22.30/0.56 & 22.05/0.56 & 24.95/0.65 \\
        
        & Noise2Void~\cite{krull_noise2void_2019} & 22.13/0.60 & 20.47/0.36 & 20.55/0.35 & 24.06/0.68 & 23.70/0.66 & 25.08/0.66\\
        & Liu et al.~\cite{liu_classification_2020} & 21.04/0.52 & 25.96/0.74 & 27.17/0.82 & 27.11/0.80 & 26.83/0.77 & 27.52/0.83\\
        & NoiseBreaker (Ours) & \textbf{23.68/0.68} & \textbf{26.33/0.82} & \textbf{27.19/0.84} & \textbf{29.94/0.90} & \textbf{29.70/0.91} & \textbf{30.85/0.92}\\
    \hline

        \multirow{5}{*}{BSD68  RGB}  & Noisy& 11.71/0.18 & 16.98/0.36 & 18.05/0.40 & 13.00/0.24 & 13.01/0.24 & 25.15/0.74 \\
        & BM3D& 21.24/0.57 & 24.72/0.66 & 24.88/0.66 & 21.96/0.59 & 22.00/0.59 & 25.73/0.70 \\
        & Noise2Void & 13.34/0.17 & 17.60/0.31 & 18.30/0.34 & 15.45/0.24 & 15.63/0.25 & 25.27/0.66\\
        & Liu et al. & 21.02/0.60 & 23.56/0.68 & 24.15/0.69 & 18.84/0.51 & 19.23/0.53  & 20.13/0.54\\
        & NoiseBreaker (Ours) & \textbf{21.88/0.71} & \textbf{26.81/0.82} & \textbf{26.58/0.82} & \textbf{25.45/0.81} & \textbf{25.20/0.80} & \textbf{29.77/0.88} \\
         
    \hhline{========}

    \end{tabular}
    \caption{Average PSNR(dB)/SSIM results of the proposed and competing methods for grayscale (top) and RGB (bottom) denoising with the noise mixtures of Table~\ref{tab:test_mixtures_table} on BSD68. Bold value indicates the best performance.}
    \label{tab:res}
\end{table*}

From the experiments, the proposed refinement of classes is found to be the best trade-off between number of classes and denoising results. Refining the classes enables more dedicated primary denoisers. On the other hand, refinement increases the classification problem complexity as well as the number of primary denoisers to be trained. Bernoulli noise is left as an only class since refinement does not bring improvement.  


\section{Experiments}\label{sec:expe}

\subsection{Data and Experimental Settings}\label{subsec:setup}
\textbf{Noise Analysis} The noise classifier $\mathcal{C}$ is fine-tuned using a subset of ImageNet~\cite{deng_imagenet:_2009}. The first 10000 images of the ImageNet evaluation set are taken, among which 9600 serve for training, 200 for validation and 200 for evaluation. To create the classes, the images are first re-scaled to $224\times224$ to fit the fixed input shape. Images are then noised according to their destination class, described in Table~\ref{tab:primary_denoisers_NBreaker_E}. The training data (ImageNet samples) is chosen to keep a similar underlying content in the images, compared to those of the backbone pre-training. Similar content with corruption variations enable to concentrate the classification on the noise and not on the semantic content. To avoid fine-tuning with the same images as the pre-training, the ImageNet evaluation set is taken. 
The weights for the backbone initialisation, pre-trained on ImageNet, are taken from the official Keras MobilNetV2 implementation. In this version, NBreaker contains 11 classes. Thus, the second layer of the head has accordingly 11 units. The classifier is trained for 200 epochs with a batch size of 64. Optimisation is performed through an Adam optimizer with learning rate $5.10^{-5}$ and default for other parameters~\cite{kingma_adam:_2014}. The optimisation is driven by a categorical cross-entropy loss. A step scheduler halves the learning rate every 50 epochs.  

\begin{figure*}
    \centering
    \includegraphics[width=0.82\linewidth]{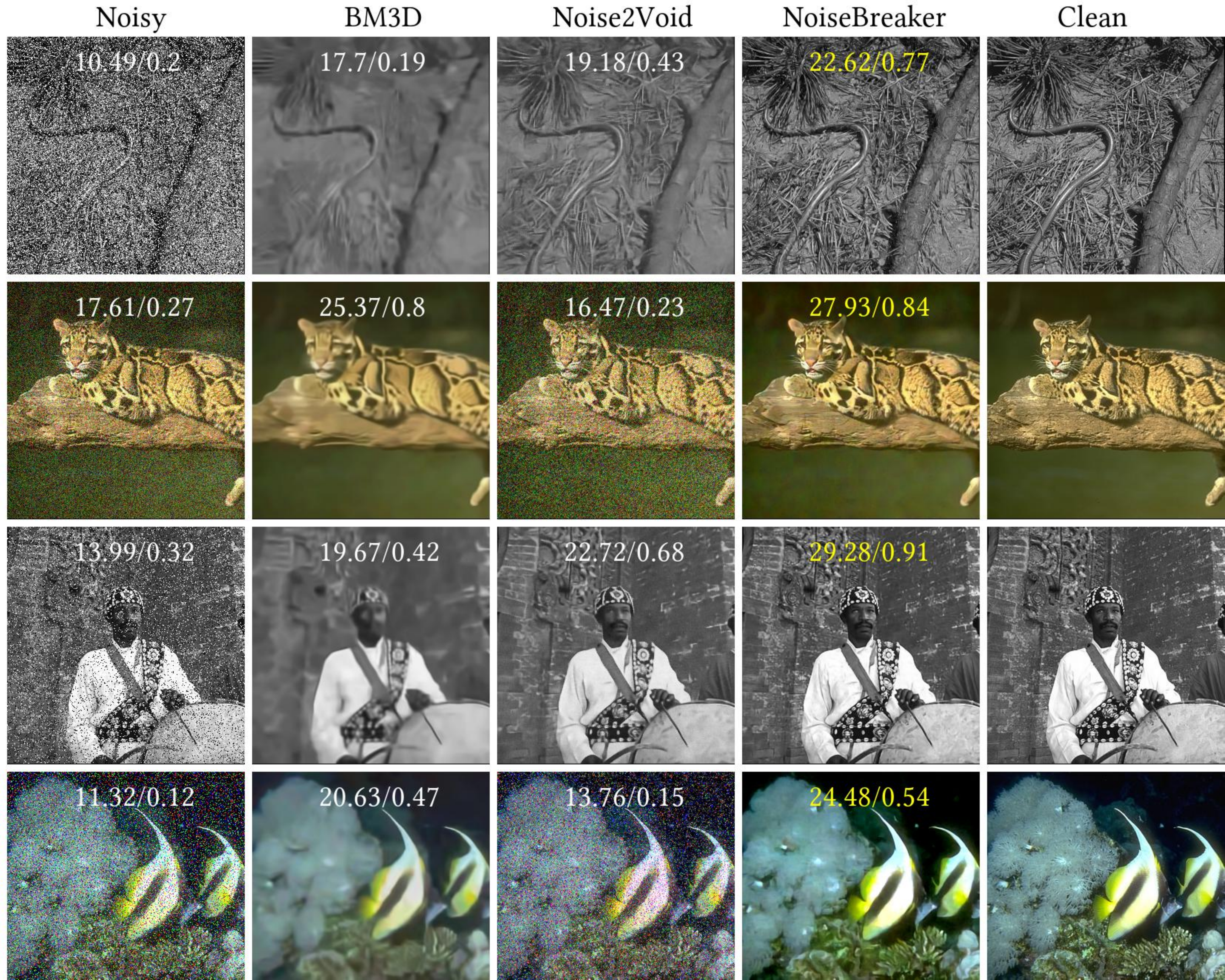}
\caption{Qualitative results for BSD68 evaluation images. Samples for first, second, third and fourth column are corrupted with mixture $C_0$, $C_2$, $C_3$ and $C_4$, respectively. \gls{psnr}/\gls{ssim} are written on each image and the better \gls{psnr} value for an image is yellow colored. \cite{liu_classification_2020} not displayed because no available code. \textbf{Better viewed on screen.}}
\label{fig:images}
\end{figure*}

\textbf{Gradual Denoising} 
For primary denoisers training, we also use the  first 10000  images  of  the  ImageNet evaluation set, re-scaled to $224\times224$.
For evaluation, the 68 images of BSD68~\cite{martin_database_2001} benchmark are used and different sequential mixtures of primary noises corrupt them. For comparison purposes, the noise types of~\cite{liu_classification_2020} are selected. Noise levels are shown in Table~\ref{tab:test_mixtures_table}. The primary noises are either \gls{awgn} with $\sigma_g\in[0,55]$,  Bernoulli noise with $p\in[0,0.4]$, white Speckle noise with $\sigma_s\in[0,55]$ or Poisson noise. $\sigma_g$, $\sigma_s$ and $p$ are randomly picked, which gives the noise distribution for the whole image. This random draw is used to prove the adaptability of our method to variable noise levels. The size of BSD68 samples is either $321\times481$ or $481\times321$. 
When evaluating the gradual denoising, $\mathcal{C}$ predicts the noise class using a patch of size $224\times224$ cropped from the image to be denoised.
The training of $\mathcal{G}$ comes down to the training of its primary denoisers. NBreaker uses only off-the-shelf architectures for primary denoisers (see table~\ref{tab:primary_denoisers_NBreaker_E}). These denoisers are trained with the parameters mentioned in their original papers. Only the training data differ since it is made of the related primary noise (according to Table~\ref{tab:primary_denoisers_NBreaker_E}).

\textbf{Compared methods} 
NoiseBreaker is compared to \gls{n2v}~\cite{krull_noise2void_2019} and \gls{bm3d}~\cite{dabov_image_2007} considered as reference methods in respectively blind and expert non-trained denoising. The recent classification-based denoiser published in~\cite{liu_classification_2020} is also considered. \gls{g2g}~\cite{cha_gan2gan:_2019} is not compared, as \gls{g2g} performances are published on non-comparable mixture noises and no code is publicly available yet. 
\gls{bm3d} is not a trained method but requires the parameter $\sigma$, the standard deviation of the noise distribution. $\sigma = 50$ is chosen since it performs better experimentally over the range of mixture noises used for evaluation. 
For \gls{n2v}, the training is carried out with the publicly available code, the original paper strategy and the data is corrupted with the previously presented synthetic evaluation mixture noise. 
For~\cite{liu_classification_2020}, results are extracted from the publication tables, as no code is publicly available.
The comparison is done based on values of \gls{ssim}, and \gls{psnr} in dB, and shown in Table~\ref{tab:res}. Qualitative results are displayed in Figure~\ref{fig:images}.

\subsection{Results}\label{subsec:res}
\begin{figure}
\centering
\subfloat[NBreaker-Grayscale]{\label{fig:conf_mat_N}{\includegraphics[width=0.48\linewidth]{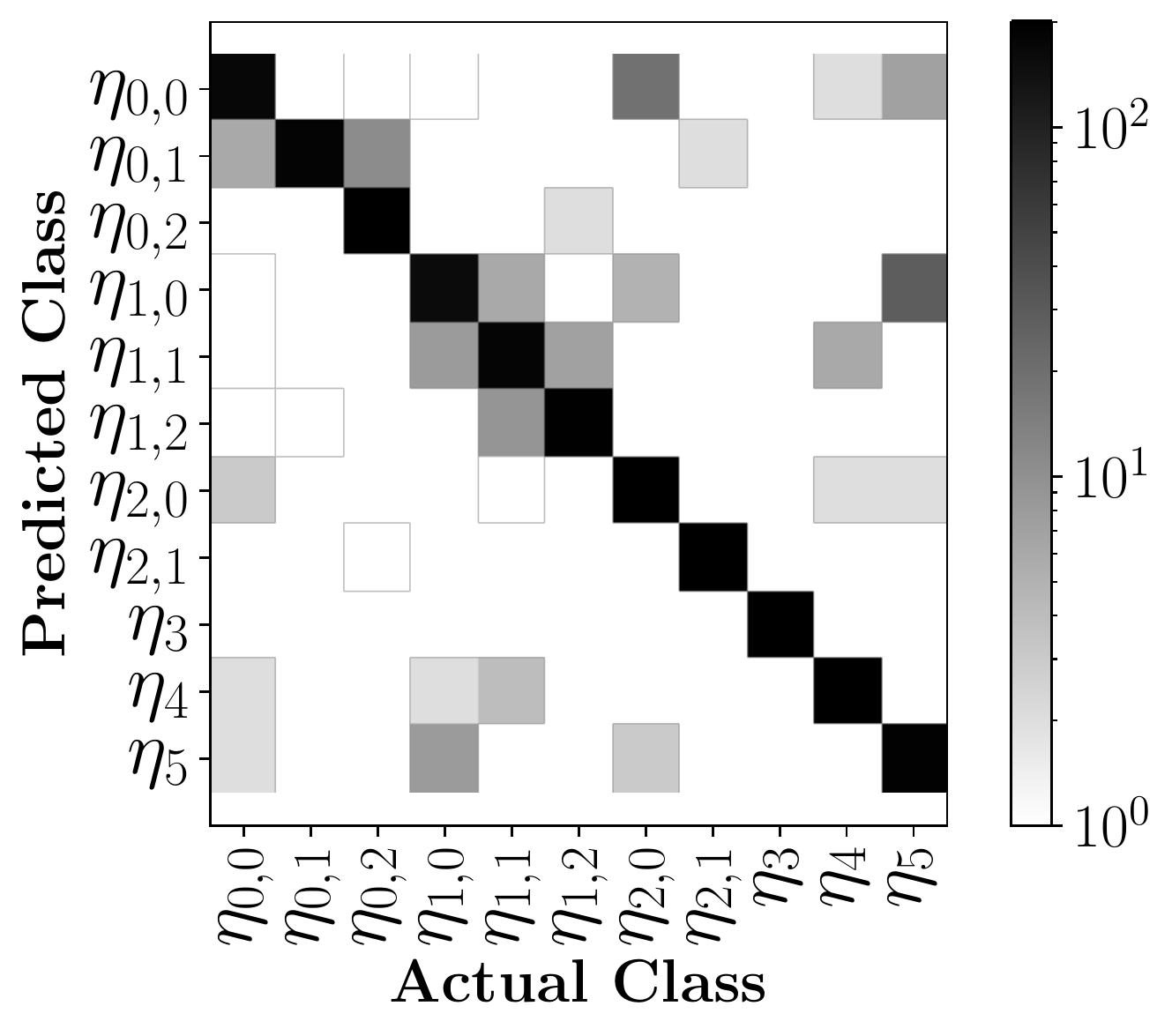}}}\hfill
\subfloat[NBreaker-RGB]{\label{fig:conf_mat_E}{\includegraphics[width=0.48\linewidth]{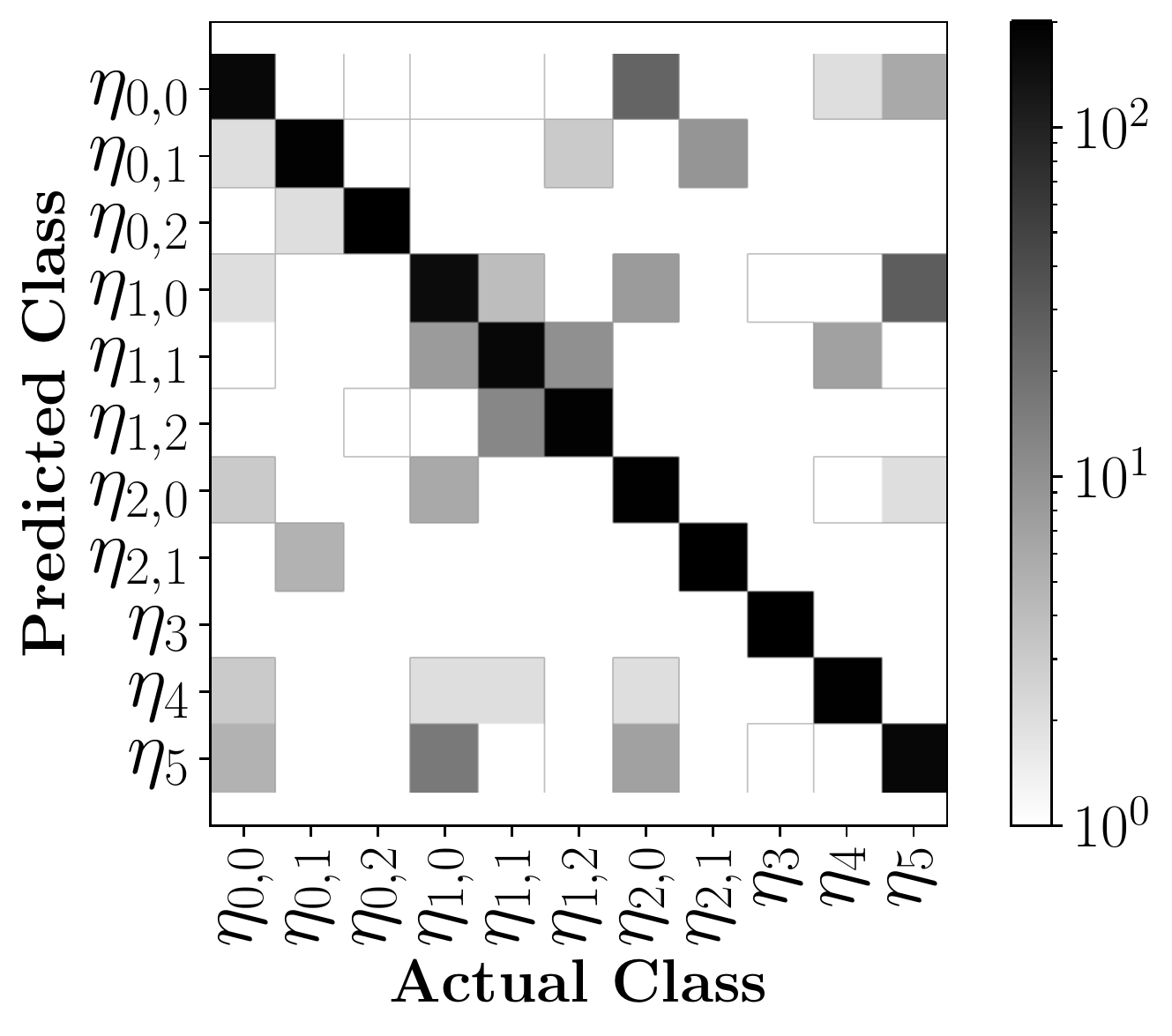}}}
\caption{Confusion matrices of noise classification evaluation in log scale. Classes content is described in Table~\ref{tab:primary_denoisers_NBreaker_E}. (a) and (b) are the results for grayscale and RGB classification, respectively.}
\label{fig:confusions_mats}
\end{figure}

\textbf{Noise Analysis}
Figure~\ref{fig:confusions_mats} presents the results of NBreaker classifiers through confusion matrices in log scale. The evaluation on 2200 images unseen during training (200 for each class) gives an accuracy score of $0.93$ for grayscale images and $0.91$ for RGB images.


The most recurrent error ($29\%$ of all the errors for grayscale, $41\%$ for RGB) is the misclassification of low noise intensity images, classified as clean ($\eta_5$) or as other low intensity noise ($\eta_{0,0}$, $\eta_{1,0}$, $\eta_{2,0}$). These effects are observable in Figure~\ref{fig:confusions_mats} (a) and (b) at ($\eta_{0,0}$, $\eta_{1,0}$), ($\eta_{2,0}$, $\eta_{0,0}$) or ($\eta_{1,0}$, $\eta_5$), where the first and second indexes represent the actual class and the predicted class, respectively. Clean images are sometimes classified as having low intensity noise ($26\%$ of all the errors for grayscale, $21\%$ for RGB). Such errors can be seen at ($\eta_5$, $\eta_{0,0}$) and ($\eta_5$, $\eta_{1,0}$). Confusions also occur between different noise levels within a unique noise type, e.g. ($\eta_{1,1}$, $\eta_{1,2}$). They represent $33\%$ of all errors for grayscale and $22\%$ for RGB. These latter errors have low impact on denoising efficiency. Indeed, this type of misclassification is caused by a noise level at the edge between two classes. As a consequence, the selected denoiser is not optimal for the actual noise level, but addresses the correct noise type. Next paragraph evaluates the performance of the classification when associated to the gradual denoiser.


\textbf{Gradual Denoising}
Table~\ref{tab:res} compares the denoising performances between \gls{bm3d}, \gls{n2v}, \cite{liu_classification_2020} and NBreaker, for the noise mixtures of Table~\ref{tab:test_mixtures_table}. Scores for noisy input images are given as baseline. 

When evaluating the methods on grayscale samples, Nbreaker operates, on average over the 6 mixtures, $2dB$ higher than the competing method of ~\cite{liu_classification_2020}. \gls{bm3d} and \gls{n2v} suffer from being applied to mixture noises far from Gaussian distributions and show average \glspl{psnr} $5dB$ under Nbreaker. The trend is the same for \gls{ssim} scores. NBreaker leads with a score of $0.85$, $13\%$ higher than \cite{liu_classification_2020} and $54\%$ higher than \gls{bm3d} and \gls{n2v}.

A first observation on RGB denoising is that \gls{n2v} underperforms, using the code and recommendations made available by the authors. Another observation is that for $C_5$, authors of~\cite{liu_classification_2020} give, in their paper, a score $5dB$ under the \gls{psnr} of noisy samples. NoiseBreaker with an average \gls{psnr} of $25.95dB$ over the 6 mixtures operates $4.8dB$ higher than \cite{liu_classification_2020}. In terms of \gls{ssim}, NBreaker shows an average score of $0.81$, an $38\%$ increase over \cite{liu_classification_2020}. 
Figure~\ref{fig:images} shows subjective results both for grayscale and RGB samples. 
NBreaker produces visually valid samples with relatively low noise levels, clear edges and contrasts. 

As a conclusion of experiments, the NoiseBreaker gradual image denoising method outperforms, on a mixture of varied primary noises, the state of the art \gls{n2v} blind method, as well as the classification-based denoiser of Liu~et~al.~\cite{liu_classification_2020}. These results demonstrate that training independently different denoisers and combining them in the NoiseBreaker iterative structure performs efficiently. 




\section{Conclusion}\label{sec:conclu}
We have introduced a gradual image denoising strategy called NoiseBreaker. NoiseBreaker iteratively detects the dominating noise with an accuracy of $0.93$ and $0.91$ for grayscale and RGB samples, respectively. Under the assumption of grayscale sequential mixture noise, NoiseBreaker operates $2dB$ over \cite{liu_classification_2020}, the current reference in gradual denoising, and $5dB$ over the state of the art blind denoiser Noise2Void. When using RGB samples, NoiseBreaker operates $5dB$ over \cite{liu_classification_2020}. 
Moreover, we demonstrate that making noise analysis guide the denoising is not only efficient on noise type but also on noise intensity. 

Additionally to the denoising performance, NoiseBreaker produces a list of the different encountered primary noises, as well as their strengths. This list makes it possible to progressively update a denoiser to a real-world noise by adding new classes of primary noise. Our future work on NoiseBreaker will include the application to a real-world eavesdropped signal denoising~\cite{lemarchand_electro-magnetic_2020}, where the corruption results from a strong sequential mixture. 

\bibliographystyle{IEEEtran}
\bibliography{refs}

\end{document}